# Synthetic Dimensions

Novel geometries can be created by coupling internal states of atoms or molecules to mimic movement in real-space

Kaden R. A. Hazzard[1, a)] and Bryce Gadway[2, b)]

[1)]*Kaden Hazzard is a theoretical physicist at Rice University in the Department of Physics and Astronomy.*
[2)]*Bryce Gadway is an experimental physicist at the University of Illinois Urbana-Champaign Department of Physics.*

(Dated: 5 June 2023)

One of the most basic laws of nature is that spatial motion is restricted to three dimensions, but a wide range of experiments have recently manipulated atoms, molecules, and light to engineer artificial matter that is so configurable that even this fundamental rule can be broken. This matter can behave as if it were in four or more spatial dimensions, or restricted to two or one dimension, as determined by the experimental design. These techniques can control not only dimensionality, but spatial geometries and potential energy landscapes. Frequently, the synthetic dimensions are created in quantum mechanical systems, so these experiments provide powerful windows into the hard-to-understand world of interacting quantum matter, which underpins many fields from quantum gravity to solid-state physics.

To understand synthetic dimensions, it helps to abstract away many details and note that physical theories – whether classical or quantum mechanical – have two ingredients: first, a set of states the system can occupy, and, second, rules for how to move between these states. For example, in classical mechanics, the state is the positions (and velocities) of particles and the rules are Newton's laws. Dimensionality is defined by these rules. Particles in one dimensional systems can step forward or backward, much like a walker on a tightrope. In three dimensions, they can also move up or down, and left or right.

The idea of synthetic dimensions is simply to allow some set of states to play the role of spatial positions, and apply controls, usually electromagnetic fields, that implement the desired rules of motion. This creates a system that is mathematically equivalent to a particle moving in a new spatial dimension, and provides novel capabilities to control geometry and other important aspects of motion in the synthetic dimension. We now look at a concrete example.

## HIGHLY EXCITED (RYDBERG) ATOMS AS A PLATFORM FOR SYNTHETIC DIMENSIONS

Synthetic dimensions have been created in numerous systems. To begin concretely, we concentrate on how synthetic dimensions have been realized using highly excited atoms known as Rydberg atoms.

This platform has been experimentally demonstrated in recent work by Profs. Tom Killian and Barry Dunning that one of us (K.H.) collaborated on as a theorist, and is being actively explored in B.G.'s lab, as well as in others around the world. Synthetic "positions" are realized by excited electronic states, while microwave-frequency electromagnetic radiation drives transitions between the electronic states, providing the "rules of the road" for what motion is allowed. These rules are mathematically equivalent to those obeyed by particles in a real lattice, for example electrons in a crystal or a molecule.

Figure 1 shows an example of how motion in a real molecule – here polyacetylene, $(C_2H_2)_n$ – can be equivalent to motion in a synthetic dimension. In the portion of the molecule shown, an electron can occupy one of six carbon atoms (dark gray balls) along the backbone, labeled as $r = 0, 1, \ldots, 5$. The equivalent synthetic positions are six highly excited electronic states, known as Rydberg states. You can think of these states essentially

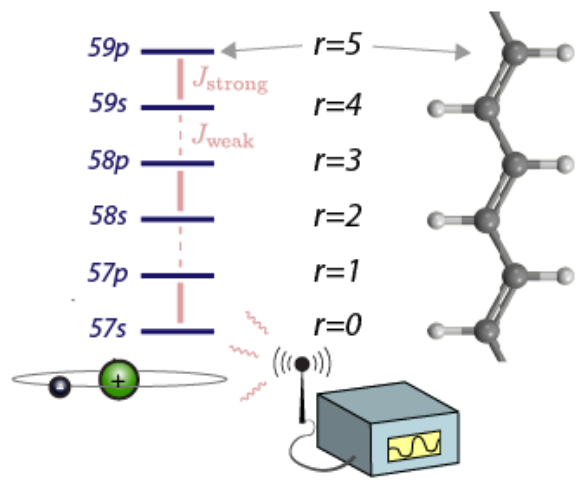

FIG. 1. Rydberg atom synthetic dimensions. Left: Highly excited electronic states of $^{84}$Sr atoms that are cooled to nanoKelvin temperatures act as position states in a synthetic dimension, while microwave electromagnetic radiation drives transitions that act as quantum tunnelings between sites in the synthetic dimension. In the illustrated scenario, states of the Rydberg atoms are connected in a one-dimensional geometry, and a staggered pattern of tunneling strengths mimics the lattice structure found in the organic conductor polyacetylene, which gives rise to interesting topological phenomena.

[a)]Electronic mail: kaden.hazzard@gmail.com
[b)]Electronic mail: bgadway@illinois.edu

as those of the textbook example of the hydrogen atom. This is because the electron in the atom is excited so far from the nucleus that it effectively interacts with the nucleus and inner electrons as if they were a single point charge. Therefore we can label the states as we would for hydrogen, with a principal quantum number $n$ and an angular momentum $\ell$. In the recent experiments, the Rydberg states were $n = 57, \ldots, 59$ angular momentum $\ell = 0, 1$ (*i.e.*, *s* and *p*) states. That establishes the equivalence of the state spaces, leaving just the rules of motion. In polyacetylene, electrons move between adjacent carbon atoms by quantum mechanical tunneling. This occurs at rates $J_{\text{strong}}$ on the double bonds and $J_{\text{weak}}$ on the single bonds. Because atoms joined by double bonds are spaced closer than ones joined by single bonds, tunneling is faster on the former, and thus $J_{\text{strong}} > J_{\text{weak}}$. The equivalent processes in the synthetic dimension are created by microwave radiation driving transitions between "adjacent" Rydberg states. Importantly, because each of the transitions between electronic levels is resonant with a different frequency, each tunneling rate can be controlled independently. The full lattice structure is formed by driving all of these transitions at the same time.

The experiments used this novel platform to observe an exciting phenomenon known as topological edge states. The polyacetylene-inspired model is known as the Su-Schrieffer-Heeger (SSH) model. The SSH model harbors spatially localized states at the boundaries of the chain, not in itself an exotic effect. Remarkably, however, the energies of these edge states are robust to all local perturbations obeying a special symmetry of the model, a phenomenon that has its roots in topology. This "topological protection" has made topology of great interest to condensed matter, materials science, and quantum information. As it turns out, topology has also been one of the most natural topics to explore in synthetic dimensions. This stems naturally from the unique abilities for designing topological configurations of the tunneling rates and potential energy landscapes, and for creating sharp system boundaries. At the same time, it is generally straightforward to perform local (*i.e.* state-resolved) measurements in synthetic dimensions. In the study by Killian and Dunning, all of these ingredients were combined to engineer protected boundary states, to directly image them through a kind of single-site-resolved "synthetic microscope," and to probe their topological robustness to controllable added disorder.

**ONGOING RESEARCH AND THE FUTURE OF SYNTHETIC DIMENSIONS**

The synthetic dimension concept is very broad, and it has been employed in ground state atoms, Rydberg atoms, molecules, and a variety of architectures using light.

While internal states, like the described Rydberg lev-els, often serve as the synthetic positions, synthetic dimensions can also be studied using motional states or arrival times of particles as the synthetic states. Some of the most notable achievements include the direct study of protected edge states and transport in topological models for neutral atoms and photons. Indeed, the capabilities to create rich tunneling arrangements and sharp edges, and to achieve local detection required for these kinds of studies follows naturally from the way that motion is induced through electromagnetic driving fields, a key feature that is common to nearly all synthetic dimensions experiments. Beyond these common capabilities, synthetic dimensions platforms have shown complementary strengths and limitations: *e.g.*, photons can have near arbitrary numbers of states but are hard to make interact strongly with one another, whereas Rydberg atoms have strong interactions but are harder to handle when coupling many states.

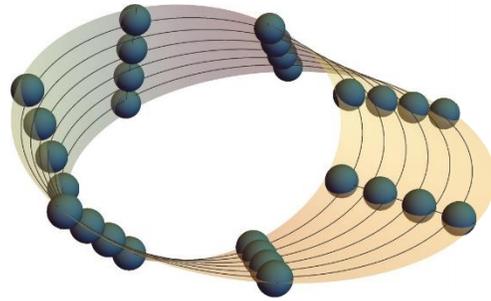

FIG. 2. Synthetic dimensions allow experiments to realize geometries that are difficult or impossible to realize in real space. For example, the famous Mobius strip geometry can be realized using Rydberg energy levels represented by blue balls and microwave couplings represented by thin lines.

Going forward, we expect this area to continue to grow, with improved coherence, larger numbers of states, and the addition of new capabilities for control, allowing new phenomena such as more complicated geometries to be explored (Fig. 2). While a large majority of experiments in this area have so far concentrated on single-particle physics or nearly non-interacting particles, the deepest mysteries about quantum matter occur for systems of interacting particles. It turns out that these are exceedingly difficult to computationally simulate in general: the computational cost to simulate even simple models grows exponentially with the number of particles, in contrast to classical systems. Here, synthetic dimensions offer a potent tool as quantum simulators: one can engineer designer systems governed by the same physics as a target system – be it the electrons in a material or quantum matter near a black hole – and explore its physics in these experiments. We expect synthetic dimension experiments to soon explore and answer important questions about such interacting quantum matter, and hopefully to raise new questions altogether.